\documentclass[12pt,a4paper]{article}

\usepackage[latin1]{inputenc}
\usepackage[english]{babel}

\usepackage{amsmath,amssymb,amsfonts}
\usepackage{graphicx}
\usepackage{epsfig}

\usepackage{a4wide}

\usepackage{booktabs,dcolumn,multirow}
\newcolumntype{d}[1]{D{.}{.}{#1}}

\usepackage{units}
\makeatletter
\def\fmslash{\@ifnextchar[{\fmsl@sh}{\fmsl@sh[0mu]}}
\def\fmsl@sh[#1]#2{%
  \mathchoice
    {\@fmsl@sh\displaystyle{#1}{#2}}%
    {\@fmsl@sh\textstyle{#1}{#2}}%
    {\@fmsl@sh\scriptstyle{#1}{#2}}%
    {\@fmsl@sh\scriptscriptstyle{#1}{#2}}}
\def\@fmsl@sh#1#2#3{\m@th\ooalign{$\hfil#1\mkern#2/\hfil$\crcr$#1#3$}}
\makeatother

\begin{document}
\begin{titlepage}
\begin{flushright}
     SI-HEP-2007-16 \\[0.2cm]
     \today
\end{flushright}

\vspace{1.2cm}
\begin{center}
     \Large\bf\boldmath
          Complete Michel Parameter Analysis \\  of the Inclusive Semileptonic $b \to c$ Transition
     \unboldmath
 \end{center}

\vspace{0.5cm}
\begin{center}
     {\sc Benjamin Dassinger, Robert Feger, Thomas  Mannel} \\[0.1cm]
     {\sf Theoretische Physik 1, Fachbereich Physik,
     Universit\"at Siegen\\ D-57068 Siegen, Germany}
\end{center}

\vspace{0.8cm}

\vspace{0.8cm}
\begin{abstract}
     \vspace{0.2cm}\noindent
     We perform a complete ``Michel parameter'' analysis of all possible helicity structures which can appear in
     the process $B \to X_c\, \ell\, \bar{\nu}_\ell$.  We take into account the full set of operators parametrizing the
     effective Hamiltonian and include the complete one-loop QCD corrections as well as the non-perturbative
     contributions. The moments of the leptonic energy as well as the combined moments of the hadronic energy
     and hadronic invariant mass are calculated including the non-standard contributions.
\end{abstract}

\end{titlepage}

\section{Introduction}
\enlargethispage{5mm}
The experimental and theoretical developments in heavy flavour physics
allow us to perform a high precision test of the flavour sector. In particular,
the enormous amount of data for semileptonic $B$ decays in combination
with very reliable theoretical methods has opened the road for a precision
determination of the CKM matrix elements $V_\text{cb}$ and $V_\text{ub}$, which are
known at a relative precision of roughly 2\,\% and 10\,\% \cite{NeubertLP07}.

Aside from testing and extracting its parameters as
precisely as possible, a second major goal of heavy flavour physics is to
look for possible effects beyond the standard model. It is genenerally
believed that flavour changing neutral currents are a good place to search for
effects of new physics, since these decays are usually loop-induced and
hence sensitive to virtual effects from high-mass states. Thus one expects
here possibly an effect which is sizable compared to the standard-model
contribution.

Semileptonic processes are tree level processes in the standard model
and thus the relative effects from new-physics contributions are likely to be
small. However, a possible right-handed admixture to the hadronic current is
completely absent in the standard model and hence such an effect would be
a clear signal for physics beyond the standard model.

In a recent publication \cite{DFM1} we considered a ``Michel parameter analysis'' \cite{Michel} of
semileptonic $B$ decays, where we considered mainly a possible right-handed
contribution to the hadronic $b \to c $ current. In the present paper we
complete the analysis of \cite{DFM1} by extending the analysis to all
possible two-quark-two-lepton operators.

There is an extensive literature on a possible non-standard model contributions to semileptonic $B$ decays \cite{GrossmanLigeti94, GrossmanLigeti95, Voloshin, Rizzo, Nierste}. However, the analysis presented here is different in two respects. First of all, our analysis is completely model-independent; however, we neglect the lepton masses and hence our analysis would need to be extended straightforwardly to include e.g. a discussion of a charged Higgs contribution as in \cite{GrossmanLigeti94, GrossmanLigeti95, Nierste}. Secondly, we consider different observables (i.\,e.\ the moments of spectra) which have become available only recently through the precise data of the $B$ factories; in this way a much better sensitivity to a non-standard contribution is expected.

In the next section we perform an effective-field-theory analysis of possible
new-physics contribution, which is kept completely model independent.
It turns out that only very few operators contribute to the semileptonic $b \to c$
transition. Compared to the usual Michel-parameter analysis, well known from
muon decay, this effective theory analysis also yields order-of-magnitude
estimates of the various contributions.

Based on these effective interactions we recompute the spectra of inclusive semi-leptonic
$b \to c $ transitons including the new interactons. We make use of the standard
heavy quark expansion (HQE) and include  QCD radiative corrections as well as
nonperturbative contributions.

In  section~\ref{sec:HQE}  we shall perform the HQE including the new-physics operators.
In subsection~\ref{sec:QCDRadCorr}  we compute the QCD radiative corrections
for the various helicity combinations of the hadronic current. We adopt the kinetic
scheme as it has been used for the calculation of semileptonic moments  in
\cite{GambinoUraltsev} and perform the complete one-loop calculation for the
new-physics terms. We note that the standard-model calculations have been performed
already to order $\alpha_s^2$  in \cite{Czarnecki}.

Subsection~\ref{sec:NonPert} we calculate the nonperturbative contributions of the
new physics  operators to order $1/m_b^2$. Finally, in section~\ref{sec:Res} we quote our
results for the various moments which are frequently used in the analysis of semileptonic
decays and conclude.

\section{Effective-Field-Theory Analysis of
   \boldmath $b \to c \ell \bar{\nu}_\ell$ \unboldmath}

It is well known that any contribution to the effective Lagrangian of some yet
unknown physics at a high scale $\Lambda$ can be written as contributions
of operators with dimensions larger than four. These operators are
$SU(3) \times SU(2) \times U(1)$ invariant and suppressed by an appropriate
power of $1/\Lambda$. We note that for such an analysis we have to make an
assumption of the yet not established Higgs Sector: We shall stick with our analysis
to the single Higgs doublet case; an extension to a type-II two-Higgs doublet as e.g.\
needed for supersymmetry is straightforward.

The lowest dimension relevant for our analysis is six; the list of relevant operators has been given in \cite{BuchmuellerWyler} and we shall use the notations of
our previous paper \cite{DFM1}. The quark and lepton fields are grouped into

\newcommand\doublet[2]{\begin{pmatrix} #1 \\ #2 \end{pmatrix}}
\begin{align}
     Q_L &= \doublet{u_L}{d_L},                  \quad
            \doublet{c_L}{s_L},                  \quad
            \doublet{t_L}{b_L}                   && \text{for the left handed quarks} \\
     Q_R &= \doublet{u_R}{d_R},                  \quad
            \doublet{c_R}{s_R},                  \quad
            \doublet{t_R}{b_R}                   && \text{for the right handed quarks} \\
     L_L &= \doublet{\nu_{e,L}}{e_L},            \quad
            \doublet{\nu_{\mu,L}}{\mu_L},        \quad
            \doublet{\nu_{\tau,L}}{\tau_L}       && \text{for the left handed leptons} \\
     L_R &= \doublet{\nu_{e,R}}{e_R},            \quad
            \doublet{\nu_{\mu,R}}{\mu_R},        \quad
            \doublet{\nu_{\tau,R}}{\tau_R}  && \text{for the right handed leptons}
\end{align}
where $Q_L$ and $L_L$ are doublets under $SU(2)_L$ and $Q_R$ and
$L_R$ are doublets under an (explicitely broken) $SU(2)_R$. Note that we also
introduced a right-handed neutrino in order to complete the right-handed lepton doubletts.

The Higgs field and its charge conjugate are
written as a $2 \times 2$ matrix
\begin{equation}
H = \frac{1}{\sqrt{2}}
     \begin{pmatrix}
          \phi_0 + i \chi_0 & \sqrt{2} \phi_+ \\
          \sqrt{2} \phi_-  & \phi_0 - i \chi_0
     \end{pmatrix}
\end{equation}
transforming under $SU(2)_L \times SU(2)_R$. The potential of the Higgs
fields leads to a vacuum expectation value (vev) for the field $\phi_0$.

The dimension-6 operators fall into two classes, the  two-quark operators with
gauge and Higgs fields and the two-quark-two-lepton operators. In our previous
analysis \cite{DFM1} we considered only the first class, and the first step towards
a full analysis is to also take into account the second class.

The list of two-quark two-lepton operators with
$SU(2)_L \times SU(2)_R$ consists of\footnote{In order to have a streamlined notation
we suppress all flavour indices in the following}
\begin{align}
     \mathcal{O}^{(i)}_{LL,LL} &= (\bar{Q}_L \Gamma_i Q_L) (L_L \Gamma_i L_L)\\
     \mathcal{P}^{(i)}_{LL,LL} &= (\bar{Q}_L \tau^a \Gamma_i Q_L) (L_L \tau^a \Gamma_i L_L)\\
     \mathcal{O}^{(i)}_{LL,RR} &= (\bar{Q}_L \Gamma_i Q_L) (L_R \Gamma_i L_R)\\
     \mathcal{O}^{(i)}_{RR,LL} &= (\bar{Q}_R \Gamma_i Q_R) (L_L \Gamma_i L_L)\\
     \mathcal{O}^{(i)}_{RR,RR} &= (\bar{Q}_R \Gamma_i Q_R) (L_R \Gamma_i L_R)\\
     \mathcal{P}^{(i)}_{RR,RR} &= (\bar{Q}_R \tau^a \Gamma_i Q_R) (L_R \tau^a \Gamma_i L_R)\\
     \intertext{while the operators with explicitely boken $SU(2)_R$ read}
     \mathcal{R}^{(i)}_{LL,RR} &= (\bar{Q}_L \Gamma_i Q_L) (L_R \Gamma_i \tau^3 L_R)  \\
     \mathcal{R}^{(i)}_{RR,LL} &= (\bar{Q}_R \Gamma_i \tau^3 Q_R) (L_L \Gamma_i L_L)    \\
     \mathcal{R}^{(i)}_{RR,RR} &= (\bar{Q}_R \Gamma_i Q_R) (L_R \Gamma_i \tau^3 L_R)  \\
     \mathcal{S}^{(i)}_{RR,RR} &= (\bar{Q}_R \tau^a \Gamma_i Q_R) (L_R \tau^a \tau^3 \Gamma_i L_R) \\
     \mathcal{T}^{(i)}_{RR,RR} &= (\bar{Q}_R \tau^a \tau^3 \Gamma_i Q_R) (L_R \tau^a \tau^3 \Gamma_i L_R)
\end{align}
Here we have defined
\begin{equation}
\Gamma_i \otimes \Gamma_i = 1 \otimes 1, \, \gamma_\mu \otimes \gamma^\mu ,
\,  \gamma_\mu \gamma_5 \otimes \gamma_5 \gamma^\mu , \,
\sigma_{\mu \nu} \otimes \sigma^{\mu \nu}
\end{equation}
Note that these operators are not all independent. Furthermore, note that
operators with the helicity combinations such as $(LR)(LR)$ cannot appear
at the level of dimension six operators, since additional Higgs fields are required
for the helicity flip.

We shall assume that the right handed neutrino acquires a large majorana
mass in which case it can be integrated out at some high scale, which we assume to
lie well above $\Lambda$. In this case $SU(2)_R$ is ``maximally broken'' for the
leptons, which means that the possible operators always have a projection
$P_- = (1-\tau^3)/2$ and thus only right handed interactions involving the
right handed charged leptons remain.

For the case at hand we are interested in the charged current interactions
containing a $b \to c $ transition. Since we eliminated the right-handed neutrino
and helicities are conserved for both currents we end up with the conclusion
that the charged leptonic current has to be left handed. Thus we have only
the operators
\begin{equation}
     O_1 = (\bar{b}_L \gamma_\mu c_L) (\bar{\nu}_{\ell,L} \gamma^\mu \ell_L)
     \quad
     O_2 =   (\bar{b}_R \gamma_\mu c_R) (\bar{\nu}_{\ell,L} \gamma^\mu \ell_L)
\end{equation}
where $\ell = e, \mu$ or $\tau$, since any helicity changing combination
has to originate from dimension-8 operators yielding an additional suppression
of a factor $v^2 / \Lambda^2$ relative to the dimension-6 contributions.

However, as has been discussed in our previous paper, helicity violating
combinations such as $(LR)(LL)$ operators can appear from the two-quark
operators with gauge and Higgs fields. These operators induce anomalous
gauge-boson couplings which are suppressed by a
factor $v^2 / \Lambda^2$. They originate from two-quark operators, which are
(at the scale of the weak bosons)
\begin{align}
     O^{(1)}_{LL} &= \bar{Q}_L \, \fmslash{L}    \,Q_L  \label{LL1}\\
     O^{(2)}_{LL} &= \bar{Q}_L \, \fmslash{L}_3  \,Q_L  \label{LL2}
\end{align}
with
\begin{align}
     L^\mu   &= H \left(i D^\mu H\right)^\dag +  \left(i D^\mu H\right)  H^\dag \\
     L^\mu_3 &= H \tau_3 \left(i D^\mu H\right)^\dag + \left(i D^\mu H\right)\tau_3 H^\dag
\end{align}
The terms proportional to $\tau_3$ have once again been included to break the
custodial symmetry explicitely.

In the same spirit we define RR-operators
\begin{align}
     O^{(1)}_{RR} &=  \bar{Q}_R  \, \fmslash{R}                         \,Q_R  \\
     O^{(2)}_{RR} &=  \bar{Q}_R  \, \left\{ \tau_3, \fmslash{R}\right\} \,Q_R \\
     O^{(3)}_{RR} &= i\bar{Q}_R  \, \left[  \tau_3, \fmslash{R}\right]  \,Q_R \\
     O^{(4)}_{RR} &=  \bar{Q}_R  \, \tau_3 \fmslash{R} \tau_3           \,Q_R
\end{align}
with
\begin{equation}
     R^\mu = H^\dag \left(i D^\mu H\right)+  \left(i D^\mu H\right)^\dag  H
\end{equation}

Using an odd number of Higgs fields we can construct invariant LR operators.
For our analysis the relevant operators are
\begin{align}
     O^{(1)}_{LR} &= \bar{Q}_L \, \left(\sigma_{\mu\nu} B^{\mu \nu}  \right)H \,Q_R \,  + \text{h.c.}\\
     O^{(2)}_{LR} &= \bar{Q}_L \, \left(\sigma_{\mu\nu} W^{\mu\nu} \right) H  \,Q_R + \text{h.c.}\\
     O^{(3)}_{LR} &= \bar{Q}_L \, \left(iD_\mu H\right) iD^\mu  \,Q_R + \text{h.c.}
\end{align}

After spontaneous symmetry breaking the LL and RR  operators contain
anomalous quark-boson couplings of the order of magnitude
$v^2 / \Lambda^2$. For the LR operators the field strenghts of the weak bosons
appear, inducing an additional factor of a quark momentum $p$, and hence the
order of magnitude is $p \, v /\Lambda^2\sim m_q v /\Lambda^2$.

At the scale of the bottom quark theses anomalous coupling terms have the
same power counting as the two-quark two-lepton operators: Integrating out the weak
bosons, their propagator together with the gauge couplings reduce to a pointlike
interaction proportional to  $g^2 / M_W^2 = 1 / v^2$.
Combining this with the order-of-magnitude of the anomalous coupling of LL and RR
$v^2 / \Lambda^2$ we find that at the scale of the bottom mass these contributions
scale in the same way as the two-quark-two-lepton operators directly induced at
the high scale $\Lambda$. For the case of LR the additional momentum $p$ of the
quark is of the order of its mass, and hence is as well of the order $v^2 / \Lambda^2$,
possibly further suppressed by a small quark Yukawa coupling.

This conclusion may be altered in a two-higgs doublet model in the case of large
$\tan \beta$, i.e.\ of a large ratio of Higgs vacuum expectation values. After integrating
out the heavy degrees of freedom  $\tan\beta$ will play the role of a coupling constant
which then may be enhanced by a large value.  In e.\,g.\ \cite{Nierste} such a scenario
has been considered, where a sizable value of $\tan \beta$ overcomes the supression
of the factor $(m_\ell m_b)  / m_{H^+}^2$ in the amplitude; in this case also scalar
contributions to the leptonic  current have to be taken into account.

To this end, the parametrization introduced in \cite{DFM1} remains valid also in
the general case, if only dimension-6 operators are included. Thus the effective
Hamiltonian reads
\begin{equation}\label{Hamiltonian2}
     \mathcal{H}_\text{eff} = \frac{4G_F V_{cb}}{\sqrt2} J_{q,\mu} J_l^\mu,
\end{equation}
where $J_l^\mu = \bar{e}\, \gamma^\mu P_-\, \nu_e$ is the usual leptonic current
and $J_{h,\mu}$ is the generalized hadronic  $b \to c$ current  which is given by
\begin{equation} \label{EnhancedGamma}
     \begin{aligned}[t]
     J_{h,\mu}  &=   c_L \ \bar{c} \gamma_\mu P_- b
                   + c_R \ \bar{c} \gamma_\mu P_+ b
                   + g_L \ \bar{c}\,i\overleftrightarrow{D_\mu} P_- b
                   + g_R \ \bar{c}\,i\overleftrightarrow{D_\mu} P_+ b \\
            &\quad + d_L \ i \partial^\nu ( \bar{c}\, i \sigma_{\mu \nu} P_- b)
                   + d_R \ i \partial^\nu ( \bar{c}\, i \sigma_{\mu \nu} P_+ b) \, ,
     \end{aligned}
\end{equation}
where $P_\pm$ denotes the projector on positive/negative chirality and $D_\mu$ is
the QCD covariant derivative.
Note that the term proportional to $c_L$ contains the standard-model contribution
as well as a possible new-physics contribution and $c_R$ may now also contain a contribution from a two-quark-two-lepton operator induced at the high scale
$\Lambda$. The gauge part $ig_3 A_\mu^a\lambda_a/2$ of the QCD covariant derivative $D_\mu$ gives rise to a new quark-quark-gluon-boson vertex.

\section{Operator Product Expansion}
\label{sec:HQE}

The operator product expansion (OPE) for inclusive decays has become
textbook material \cite{ManoharWise}.  For the case of inclusive semileptonic
decays the OPE is formulated for the $T$-product of the two hadronic currents
\begin{equation} \label{hadcorr}
     T_{\mu \nu} = \int d^4 x \,  e^{-ix(m_b v-q)}  \langle B(p) |\bar{b}_v (x) \Gamma_\mu  c(x)
     \, \bar{c}(0) \Gamma_\nu^\dagger b_v(0)  | B(p) \rangle
\end{equation}
where $\Gamma$ is the combination of Dirac matrices and derivatives given in \eqref{EnhancedGamma},
$v = p/M_B$ is the four velocity of the decaying $B$~meson and $q$ is the momentum
transferred to the leptons. The quantity $T_{\mu \nu}$ is expanded
in inverse powers of the scale of the order $m_b$, where $m_b$ is the the heavy quark mass. Technically this procedure
is an OPE for the product of the two currents.

The standard-model calculation has been performed at tree level up to order
$1/m_b^4$ and it turns out that the non-perturbative corrections are small. The  radiative
corrections have been computed to order $\alpha_s$,
$\beta_0 \alpha_s^2$  and recently also to order $\alpha_s^2$ for the leading
(i.\,e.  the parton model) term \cite{Czarnecki} and to order $\alpha_s$
for the term of order $1/m_b^2$ involving $\mu_\pi^2$.

In the following we shall consider the perturbative and non-perturbative contributions to the
OPE, performed with the modified current (\ref{EnhancedGamma}). We shall compute the complete
one-loop contributions as well as the leading non-perturbative corrections proportional to $\mu_\pi^2$
and $\mu_G^2$.

\subsection{QCD Corrections and Renormalization Group Analysis} \label{sec:QCDRadCorr}

\begin{figure}
  \begin{center}
    \includegraphics{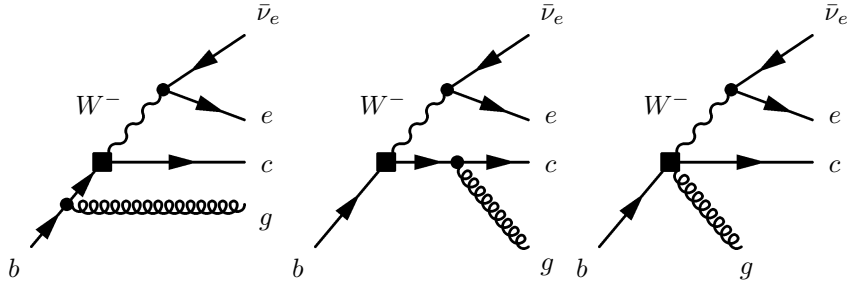}
    \caption{\label{fig:real}Real Corrections}
  \end{center}
\end{figure}
\begin{figure}
  \begin{center}
    \includegraphics{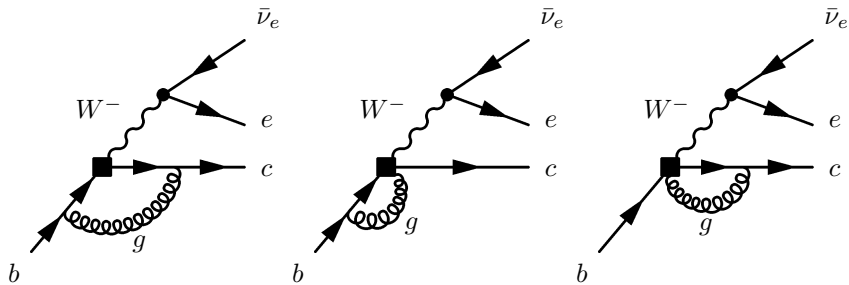}
    \caption{\label{fig:virtual}Virtual Corrections}
  \end{center}
\end{figure}

The calculation of the QCD radiative corrections has been performed in \cite{veryold2} and the results
in the kinetic scheme have been given in \cite{GambinoUraltsev} for the semileptonic moments in the
standard model. In order to perform an analysis of possible non-standard contributions we have to calculate
the QCD radiative corrections for the current \eqref{EnhancedGamma} to order $\alpha_s$. Thus we have
to evaluate the Feynman diagrams shown in fig. \ref{fig:real} and \ref{fig:virtual} for the real and virtual corrections respectively. Note that the scalar current (i.\,e. the terms proportional to $g_{L/R}$) induces new vertices shown in the Feynman diagrams at right.
The real and virtual corrections are individually IR-divergent. We regulated the IR-divergence by introducing a gluon mass which drops out upon summation of the real and virtual correction being IR-convergent.
In the calculation of  the virtual corrections the wave function renormalizations of the b and c quark field also
have to be included.

The total amplitude consists of the sum of the standard-model contribution and the one from the new-physics
operators. Since the new-physics piece is of order $1/\Lambda^2$, we shall include only the interference term
of the standard model with this contribution. The square of the new-physics term is already of order
$1/\Lambda^4$ and has to be neglected, since we compute only up to this order.  Thus we compute
\begin{equation}
     \text{d}\Gamma=\frac{1}{2m_b}\bigl(\langle cl\nu|c_L\mathcal{H}_\text{eff}^\text{SM}|b\rangle\langle cl\nu|\mathcal{H}_\text{eff}|b\rangle^* + \langle cl\nu|\mathcal{H}_\text{eff}|b\rangle\langle cl\nu|c_L\mathcal{H}_\text{eff}^\text{SM}|b\rangle^*\bigr)\text{d}\phi_\text{PS}
\end{equation}
where $\text{d}\phi_\text{PS}$ is the corresponding phase space element and
\[\mathcal{H}_\text{eff}^\text{SM}=\frac{4G_F V_{cb}}{\sqrt2}\bigl(\bar{c} \gamma_\mu P_- b\bigr)\bigl(\bar{e}\, \gamma^\mu P_-\, \nu_e\bigr)\]
is the standard-model effective Hamiltonian, which has the same helicity structure as the new-physics
contribution proportional to $c_L$.

The relevant Feynman rules for the new-physics operators at tree level can be read off from
\eqref{EnhancedGamma}; note that the terms involving $g_L$ and $g_R$  yield a boson-gluon-quark-antiquark
vertex in order to maintain QCD gauge invariance.

It is well known that the left- and right-handed currents do not have anomalous dimensions and hence the
parts of \eqref{EnhancedGamma} with $c_L$ and $c_R$ are not renormalized. However,
the scalar and tensor contributions have anomalous dimensions and hence we need to normalize these
operators at some scale and run them down to the scale of the bottom quark.

To this end, we have to calculate the anomalous-dimension matrix of these currents to set up the renormalization
group equation. It can be obtained from the requirement that the physical matrix elements must not depend
on the renormalization scale $\mu$:
\begin{equation}\label{rencondition}
   0=\frac{\text{d}}{\text{d}\ln\mu}\langle c\ell\nu_e |\mathcal{H}_\text{eff} | b \rangle
\end{equation}
Inserting the OPE for the Hamiltonian we get:
\begin{eqnarray}
\langle c\ell\nu_e |\mathcal{H}_\text{eff} | b \rangle &=& \frac{4 G_\text{F}V_\text{cb}}{\sqrt{2}}\,
\cdot\langle c\ell\nu_e | \left[  c_L (\bar{c}\,\gamma_\mu P_- b) (\bar{e}\,\gamma^\mu P_- v_e)
      + c_R (\bar{c}\,\gamma_\mu P_+ b) (\bar{e}\,\gamma^\mu P_- v_e)
\right] | b \rangle   \nonumber \\
&& + \frac{4 G_\text{F}V_\text{cb}}{\sqrt{2}}\,
\vec{C}\cdot\langle c\ell\nu_e |\vec{\mathcal{O}} | b \rangle,
\end{eqnarray}
with
\renewcommand{\arraystretch}{1.2}
\begin{equation}\label{vectors}
\vec{C}=
   \begin{pmatrix}
      g_L\\
      g_R\\
      d_L\\
      d_R\\
      c_L^{m_b}\\
      c_R^{m_b}\\
      c_L^{m_c}\\
      c_R^{m_c}\\
   \end{pmatrix}
\qquad
\vec{\mathcal{O}}=
   \begin{pmatrix}
      (\bar{c}\,i\overleftrightarrow{D_\mu} P_- b) (\bar{e}\,\gamma^\mu P_- v_e)  \\
      (\bar{c}\,i\overleftrightarrow{D_\mu} P_+ b) (\bar{e}\,\gamma^\mu P_- v_e)  \\
      (i\partial^\nu(\bar{c}\,i\sigma_{\mu\nu}P_- b)) (\bar{e}\,\gamma^\mu P_- v_e)   \\
      (i\partial^\nu(\bar{c}\,i\sigma_{\mu\nu}P_+ b)) (\bar{e}\,\gamma^\mu P_- v_e)  \\
      (m_b\, \bar{c}\,\gamma_\mu P_- b) (\bar{e}\,\gamma^\mu P_- v_e) \\
      (m_b\, \bar{c}\,\gamma_\mu P_+ b) (\bar{e}\,\gamma^\mu P_- v_e) \\
      (m_c\, \bar{c}\,\gamma_\mu P_- b) (\bar{e}\,\gamma^\mu P_- v_e) \\
      (m_c\, \bar{c}\,\gamma_\mu P_+ b)(\bar{e}\,\gamma^\mu P_- v_e)  \\
   \end{pmatrix}.
\end{equation}
\renewcommand{\arraystretch}{1}
where the operators $ \vec{{\cal O}} $ are of dimension seven.

In the following we consider the renormalization group mixing of these dimension-four operators. The
calculation of the one-loop anomalous dimension is standard. We define the anomalous dimension
matrix $\gamma$ by:
\begin{equation}\label{GAMMA}
  \frac{\text{d}\vec{C}}{\text{d}\ln\mu}=\gamma^T(\mu)\,\vec{C}
\end{equation}
and compute $\gamma$ from the divergencies of the renormalization constants in the usual way.
We obtain
\begin{equation}
\gamma^T(\mu)=\frac{2\alpha_s(\mu)}{3\pi}
       \begin{pmatrix}
          0 & 0 & 0 & 0 & 0 & 0 & 0 & 0\\
          0 & 0 & 0 & 0 & 0 & 0 & 0 & 0\\
          \textbf{1} & 0 & \textbf{1} & 0 & 0 & 0 & 0 & 0\\
          0 & \textbf{1} & 0 & \textbf{1} & 0 & 0 & 0 & 0\\
          0 & \textbf{3} & 0 & 0 & \textbf{3}  & 0 & 0 & 0\\
          \textbf{3} & 0 & 0 & 0 & 0 & \textbf{3}  & 0 & 0\\
          \textbf{3} & 0 & 0 & 0 & 0 & 0 & \textbf{3} & 0\\
          0 & \textbf{3} & 0 & 0 & 0 & 0 & 0 & \textbf{3}\\
      \end{pmatrix}
\end{equation}
The renormalization group equation for the Wilson coefficient is
\begin{equation}\label{RGE}
  \biggl(\frac{\partial}{\partial\ln\mu}+\beta(\alpha_s)\frac{\partial}{\partial\alpha_s}\biggr)\,\vec{C}=\gamma^T\bigl(\alpha_s(\mu)\bigr)\,\vec{C}.
\end{equation}
We seek a solution of this equation with the initial conditions
\begin{equation}
c_{L/R}^{m_b} (\Lambda) = 0 = c_{L/R}^{m_c} (\Lambda)  \, ,
\end{equation}
since the matching of the left- and right-handed currents is performed by fixing the coefficients
$c_L$ and $c_R$ and all additional contributions are only due to renormalization group running.
Inserting the one-loop expressions we obtain
\begin{equation}
     \begin{aligned}\label{running}
     c_{L/R}(\mu)&=c_{L/R}(\Lambda)\\[2mm]
     g_{L/R}(\mu)&=g_{L/R}(\Lambda)\\
     d_{L/R}(\mu )&= \bigl(g_{L/R}(\Lambda)+ d_{L/R}
     (\Lambda )\bigr)\left(\frac{\alpha_s(\Lambda)}{\alpha_s(\mu
     )}\right)^{\frac{4}{3 \beta_0}}-g_{L/R}(\Lambda) \\
     c_{L/R}^{m_b}(\mu )&= g_{R/L}(\Lambda) \left(\left(\frac{\alpha_s(\Lambda
     )}{\alpha_s(\mu )}\right)^\frac{4}{\beta_0}-1\right)  \\
     c_{L/R}^{m_c}(\mu )&= g_{L/R}(\Lambda) \left(\left(\frac{\alpha_s(\Lambda
     )}{\alpha_s(\mu )}\right)^\frac{4}{\beta_0}-1\right) \\
     \end{aligned}
\end{equation}
One may reexpand \eqref{running} using the one-loop  expression for the strong coupling constant
and obtain the logarithmic terms of the one-loop calculation.  However, the straight-forward
one-loop calculation also yields finite terms, which depend on the choice of the renormalization
scale $\mu$. It is well known that in order to fix this dependence on the renormalization scale, one
would need to include the running at two loops, which, however, goes beyond the scope of the present
paper.  Rather we shall fix this scale to be $\mu = m_b$, which is the relevant scale of the decay process,
assuming  that  the full NLO calculation would fix a scale of this order.

The advantage of this procedure is that the kinematic effects, which lead to a distortion of the spectra and
thus have an impact on the moments, are given by these finite terms. We expect that a full NLO calculation
will lead to very similar results.

\subsection{Mass Scheme}

The calculation of the process is usually set up with pole masses of the particles, which is known not to be
a well defined mass. The problems manifest themselves by abnormally large radiative corrections when
the pole mass scheme is used. It has been discussed extensively in the literature that an appropriately
defined short-distance mass is better suited for the OPE calculation of an inclusive  semileptonic rate.

In the present analysis we will use the kinetic mass scheme, where the mass is defined by a non-relativistic
sum rule for the kinetic energy \cite{UraltsevSR}.  At one-loop level the kinetic mass is related to the pole mass by
\begin{equation}\label{massscheme}
 m_\text{q}^\text{kin}(\mu_f)=m_\text{q}^\text{pole}\Biggl[1-\frac{4}
{3}\frac{\alpha_s}{\pi}\biggl(\frac{4}{3}\frac{\mu_f}{m_b}+\frac{\mu_f^2}
{2m_\text{b}^2}\biggr)\Biggr]
\end{equation}
where $\mu_f$ is a factorization scale for removing contributions below from the mass-definition. The factorization scale is  set to \unit[1]{GeV} since this is the typical energy release in the process. This low renormalization scale is
in fact the reason why the $\overline{MS}$ scheme is inappropriate.

The ratio $\rho=m_\text{c}^2/m_\text{b}^2$ is rather stable under the choice of schemes (provided that the
same scheme is chosen for both $m_b$ and $m_c$) and thus  the choice of the  mass scheme enters only
through the $m_\text{b}^5$ dependence of the rates.  It is well known from the calculation in the standard model
that the ${\cal O}(\alpha_s) $ corrections from the relation of the kinetic mass with the pole mass
\begin{equation}
 \biggl(\frac{m_\text{q}^\text{pole}}{m_\text{q}^\text{kin}(\unit[1]
{GeV})}\biggr)^5\approx  1+2.0899\,\frac{\alpha_s}{\pi}.
\end{equation}
compensate to a large extent the radiative corrections to the rates computed in the pole scheme, leaving only
small QCD radiative corrections. It turns out that this also is the case in our calculation including the anomalous
couplings.

\subsection{Non-perturbative Corrections} \label{sec:NonPert}
The nonpertubative corrections at tree level, including the modified current  (\ref{EnhancedGamma}),
have been studied in \cite{Feger_DassingerDiplom}, however, these results have not yet been
published, and hence we shall quote  these results in the following.

\begin{figure}[h]
  \begin{center}
	\includegraphics{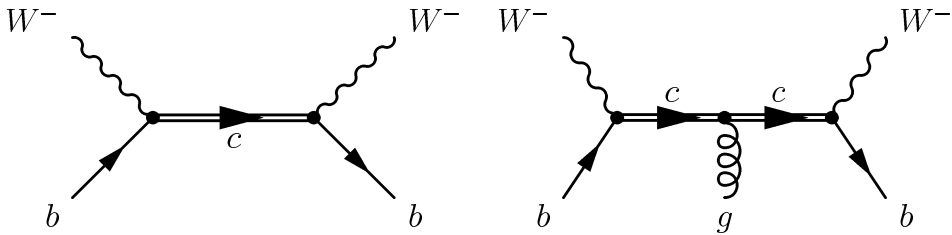}
	\caption{\label{fig:OPEtree}Nonpertubative corrections at tree level}
  \end{center}
\end{figure}

The calculation of the nonperturbative corrections at tree level requires to compute the Feynman
diagram shown in fig. \ref{fig:OPEtree}, where the one-gluon graph is needed to obtain the matching
coefficient of the chromo-magnetic moment operator.

The calculation is standard and yields somewhat lengthy results, thus we defer the presentation of
these expressions to the appendix.

To the order we calculate, the nonperturbative effects are parametrized
by the kinetic energy $\mu_\pi^2$ and the chromomagnetic moment $\mu_G^2$, which  are small
quantities compared to the $b$-quark mass.
Inserting the values extracted from $b \to c$ semileptonic decays we find
\begin{equation}
\frac{\mu_\pi^2}{m_b^2} \sim \frac{\mu_g^2}{3 \,m_b^2} \sim 0.02  \, .
\end{equation}
Hence the  non-perturbative corrections are tiny compared to the leading terms, as long as
there are no abnormally large coefficients or the leading term vanishes.
As it has been investigated in \cite{Feger_DassingerDiplom} this as well holds true for the
new-physics contributions parametrized by (\ref{EnhancedGamma}).

\section{Results and Discussion} \label{sec:Res}

We have evaluated the new-physics contributions to the various moments of the leptonic and hadronic energy
and the hadronic invariant mass spectra. We have included tree-level partonic and $1/m_b^2$ corrections as well as the
QCD radiative corrections at one-loop with a renormalization group treatment as described in the last section.
The hadronic energy and the hadronic invariant mass of the decay products can be written as
\begin{equation} \label{VarHad}
\begin{aligned}
E_{Had}&=v\cdot(p_B-q)=m_B-v\cdot q\\
s_{Had}&=(p_B-q)^2=m_B^2-2 m_B\, v\cdot q +q^2,
\end{aligned}
\end{equation}
where $m_B$ and $p_B=m_B\, v$ are the mass and the momentum of the B meson and $q$ is the momentum of the leptonic
system.  The $B$-meson mass can be expanded as
\begin{equation}
m_B = m_b + \bar\Lambda + \frac{\mu_\pi^2+\mu_g^2}{2 m_b} + \cdots .
\end{equation}
Thus it is possible to relate the hadronic variables
in (\ref{VarHad}) to the partonic ones
\begin{equation} \label{VarPart}
\begin{aligned}
\hat{E}_0 = \frac{E_0}{m_b} &=\frac{v\cdot(p_b-q)}{m_b}=1 - v\cdot \hat q\\
\hat{s}_0 = \frac{s_0}{m_b^2}&=\frac{(p_b-q)^2}{m_b^2}=1-2 v\cdot \hat q +\hat q^2,
\end{aligned}
\end{equation}
where $p_b$ is the the b quark momentum. In the following we shall quote the results in terms
of the partonic variables (\ref{VarPart}).

We have also included a cut on the charged lepton energy since such a cut has to be used in the experimental analysis.
The results may be obtained as {\tt FORTRAN} code from the authors. In order to have a qualitative discussion
of the results, we give the results in tables \ref{tab:L}-\ref{tab:Hcut} for various moments without
an energy cut for the charged lepton energy $E_l$ and with a cut of \unit[1]{GeV} for this quantity. In table \ref{tab:L}
we list the results for the moments
\begin{equation}
	L_n = \frac{1}{\Gamma_0} \int_{E_\text{cut}} d \hat E_l \, \hat E_l^n \, \frac{\text{d} \Gamma}{\text{d} \hat E_l} \, ,
\end{equation}
in table~\ref{tab:LSummed} we consider the scale dependence of the $L_n$,
and in the tables \ref{tab:Hnocut} and \ref{tab:Hcut} we quote
\[
	H_{ij} = \frac{1}{\Gamma_0}
	\int_{E_\text{cut}} \text{d} \hat E_l \, \int \text{d}\hat s_0 \, \text{d} \hat E_0
	(\hat s_0-\rho)^i \, E_0^j \,
	\frac{\text{d}^3 \Gamma}{\text{d} \hat E_0 \, \text{d}\hat s_0 \,  \text{d} \hat E_l}
\]
with $\rho=m_b^2/m_c^2$, where the normalization
\begin{equation} \label{TotRateTree}
	\Gamma_0 = \frac{G_\text{F}^2|V_{cb}|^2 m_b^5}{192 \pi^3}
	\bigl(1-8\rho -12 \rho^2 \ln \rho + 8 \rho^3 - \rho^4 \bigr)
\end{equation}
is given in terms of the partonic rate. Note that we perform the calculations in the kinetic scheme, and we
also insert the value of the kinetic mass in the normalization $\Gamma_0$.

The entries in the tables contain the coefficients corresponding to the expansion of the various moments:
\begin{align}
	L_n={}	&  c_L^2 L_n^{(c_Lc_L)} +c_L c_R L_n^{(c_Lc_R)} +c_L d_L L_n^{(c_Ld_L)} \label{LN} \\ \nonumber
	              & + c_L d_R L_n^{(c_Ld_R)} + c_Lg_L L_n^{(c_Lg_L)} + c_Lg_R L_n^{(c_Lg_R)} \\
	H_{ij}={}& c_L^2 H_{ij} ^{(c_Lc_L)} +c_L c_R H_{ij} ^{(c_Lc_R)} +c_L d_L H_{ij} ^{(c_Ld_L)}  \label{HN} \\  \nonumber
	             & +c_L d_R H_{ij}^{(c_Ld_R)} +c_L g_L H_{ij} ^{(c_Lg_L)} + c_Lg_R H_{ij} ^{(c_Lg_R)}
\end{align}
where all the coefficients have an expansion in $\alpha_s$ and in $1/m_b$
\begin{eqnarray*}
L_n^{(c_1 c_2)} &=& L_n^{(c_1 c_2;\,m_b^0,\alpha_s^0)} + \frac{\mu_\pi^2}{m_b^2} L_n^{(c_1 c_2;\,m_b^2, \alpha_s^0)} + \frac{\mu_g^2}{3 \,m_b^2} L_n^{(c_1 c_2;\,m_b^2, \alpha_s^0)} + \cdots +\frac{\alpha_s}{\pi}  L_n^{(c_1 c_2;\,m_b^0, \alpha_s^1)}  + \cdots \\
H_{ij}^{(c_1 c_2)} &=& H_{ij}^{(c_1 c_2;\,m_b^0, \alpha_s^0)} + \frac{\mu_\pi^2}{m_b^2} H_{ij}^{(c_1 c_2;\,m_b^2, \alpha_s^0)} + \frac{\mu_g^2}{3 \,m_b^2} H_{ij}^{(c_1 c_2;\,m_b^2, \alpha_s^0)}  +  \cdots +\frac{\alpha_s}{\pi}  H_{ij}^{(c_1 c_2;\,m_b^0, \alpha_s^1)}  + \cdots
\end{eqnarray*}
where we have only shown the terms which we have calculated.  The values of couplings to be inserted in
\eqref{LN} and \eqref{HN} are the ones at the high scale $\Lambda$. Furthermore, for the numerical
analysis we use $m_b^{\rm kin} (1\, \unit{GeV}) = \unit[4.6]{GeV}$ and $\rho = m_c^2 / m_b^2 = 0.0625$.

The results of the calculations are displayed in tables Tab. \ref{tab:L}-\ref{tab:HSummed} in the appendix. Table \ref{tab:L}
contains the results for the leptonic moments normalized to the total leptonic rate at tree level
(\ref{TotRateTree}) for all lepton energies and for a cut  of 1 GeV on the lepton energy.
It turns out that the radiative corrections to the scalar and tensor admixtures are sizable, i.\,e.
the $\alpha_s/\pi$ coefficients are large. In addition, these coefficients have the opposite sign
as the tree level piece, and hence a substantial reduction of the tree result is expected.

Table \ref{tab:LSummed} contains the sum of the tree level and the $\alpha_s$ contributions using
the one-loop expression for the running coupling $\alpha_s$.  As discussed above,  the full
NLO expressions for the scalar and tensor couplings are not available yet and hence a residual
scale dependence remains. We expect the scale to be of the order of $m_b$ and hence we
evaluate the expressions for $\mu = m_b / 2 $, $m_b$ and $2 m_b$. For $c_L^2$ as well as for
$c_L c_R$ the scale dependence is weak and originates from yet unknown NNLO effects.
Due to the large $\alpha_s / \pi$ coefficients the scale dependence for the
tensor couplings is sizable, while it is huge for the scalar couplings, since the tree contribution
is almost cancelled by the radiative correction. A full NLO calculation will very likely not improve
this situation and hence we have to conclude that we will not have a good sensitivity to the
tensor couplings and practically no sensitivity to the scalar couplings, at least for the lepton-energy
moments.

The coefficients of the nonperturbative contributions at tree level are in general of similar size
as the ones of the $\alpha_s$  corrections. Since $\alpha_s / \pi \sim  \mu_\pi / m_b^2 $, the non-perturbative
corrections are of similar importance as the radiative ones. However, the leptonic moments are
all dominated by the tree-level contribution and hence the radiative as well as the nonperturative
corrections to the moments are small.

Tables \ref{tab:Hnocut} and \ref{tab:Hcut} contain the various hadronic moments
computed without and with a cut on the lepton energy. For the $i = 0$ moments we have to draw the
same conclusion as for the leptonic moments: The scalar and tensor couplings have large
and opposite-sign coefficients compared to the tree level piece; this leads in the same way to
a sizable reduction of the tree level result as well as  to a large scale dependence, which is shown
in table \ref{tab:HSummed}, where the result up to order $\alpha_s$ is shown.

Clearly the moments with $i > 0$ do not have a tree level contribution at the partonic level since
the tree-level partonic rate is proportional to the mass shell delta function $\delta (\hat{s}_0 - \rho)$.
For these moments the leading contributions are at order  $\alpha_s$ or $1/m_b^2$. Hence  their
dependence on the scale is given by the dependence of $\alpha_s$. However, here the radiative
corrections are small compared to the non-perturbative ones.  The non-perturbative corrections
at tree level contain also derivatives of the mass shell delta function $\delta (\hat{s}_0 - \rho)$,
where at leading order $1/m_b$ the maximum number of derivatives is two. Due to this, the first and
second $i$ moments are of order $1/m_b^2$;  higher moments with
$i > 2$ will only have contributions of order $1/m_b^3$ or higher.

The sensitivity to a possible new-physics contribution is mainly limited by the  precision of the
standard-model calculation. Current analyses use up to
the second moments in both the leptonic energy and  the invariant mass squared. The highest moments
included in the standard-model analyses are (roughly) sensitive to terms of the order $1/m_b^3$ which is
the highest order in the $1/m_b$ expansion included in the fit. The size of this terms together with the size of the
$\alpha_s^2$ corrections may serve as a conservative estimate of the uncertainties of the standard model
calculation, which at the end determines the sensitivity to a possible new-physics contribution.
Furthermore, an inclusion of higher moments in the fit, including the new contributions,
(in particular with $ i > 2$) needs the calculation of the $1/m_b^3$ terms for the new-physics contributions.
As the impact of such hadronic mass moments to the fit is small we did not include a table of them in this paper,
but the results of the calculation can be obtained by the authors in a \textsc{Fortran} or \textsc{Mathematica} file.

\section{Summary and Conclusions}

This work completes the analysis of possible new physics effects in inclusive semileptonic
$B$ decays. Starting from a general ansatz for anomalous couplings in semileptonic decays
we compute the effects on leptonic and hadronic moments which are used in the analysis
of inclusive semileptonic decays.

As far as the leptonic moments are concerned, the QCD radiative corrections turn out
to be as important as the nonperturbative ones. We have presented the complete expressions
to order $\alpha_s$ and to order $1/m_b^2$ including the new-physics pieces.

This holds also true for the hadronic energy moments. However, the hadronic mass moments
(taken with respect to $m_c^2$) do not have a tree-level contribution. Hence the nonperturbative
corrections of order $1/m_b^2$ as well as the terms of order $\alpha_s$ are the leading
contributions in the heavy quark expansion. It turns out that, numerically, the non-perturbative
contributions are in general dominant.

For the leptonic moments and for the $i = 0$ hadronic moments the radiative correction for the scalar
and tensor couplings turn out to be sizable. This leads to a substantial reduction of the moments and
in the case of the scalar coupling to an almost cancellation between tree level and the radiative
corrections, which induces a  large scale
dependence. Hence the sensitivity to scalar and tensor couplings of the moments is limited.
However, the moments with $i \neq 0$ appear first at order $\alpha_s$ and have a resonable sensitivity
to scalar and tensor couplings.

The standard analysis in semileptonic decays is to perform a combined fit  of $V_\text{cb}$, the
quark masses and the HQE parameters, usually up to order $1/m_b^3$. We propose to use the
results given here to include the anomalous couplings induced by possible new physics into
such a fit.

The effective-theory analysis indicates that a right-handed admixture could be the largest effect.
Since the radiative corrections to the right-handed currents are completely known to NLO and the
size of the coefficients indicate a good sensitivity to the anomalous coupling $c_R$ this coefficient
should be the first to be searched for in a moment analysis.

\subsection*{Acknowledgements}
TM acknowledges helpful discussions with I. Bigi.
This work was partially supported by
the German Research Foundation (DFG) under contract No.
MA1187/10-1, and by the German Minister of Research (BMBF), contract No. 05HT6PSA.

\appendix
\section{Nonperturbative corrections to the non-standard currents}
In this appendix we show the results for the new-physics contributions based on
(\ref{EnhancedGamma}). We have calculated the interference term of the standard-model
contribution with the current (\ref{EnhancedGamma}) and list the terms proportional to the
coupling constants. We obtain 

\vspace{0.5cm}

\begin{align*}
\left(\frac{\text{d} \Gamma}{\text{d} y}\right)^{c_L c_L}&
\begin{aligned}[t]
=&\left(\frac{2 (y-3) y^2 \rho ^3}{(y-1)^3}-\frac{6 y^2 \rho ^2}{(y-1)^2}-6 y^2 \rho +2 (3-2 y) y^2\right) \\
&+\left(\frac{4 \left(y^2-5 y+10\right) \rho ^3 y^3}{3 (y-1)^5}+\frac{2 (5-2 y) \rho ^2 y^3}{(y-1)^4}+\frac{10 y^3}{3}\right) \frac{\mu_\pi^2}{m_b^2}\\
&+\left(\frac{10 y^2 \left(y^2-4 y+6\right) \rho^3}{(y-1)^4}-\frac{18 (y-2) y^2 \rho ^2}{(y-1)^3}+\frac{12 y^2 (2 y-3) \rho }{(y-1)^2}+2 y^2 (5 y+6)\right) \frac{\mu_g^2}{3 \,m_b^2}
\end{aligned}\\ \\
\left(\frac{\text{d} \Gamma}{\text{d} y}\right)^{c_L c_R}&
\begin{aligned}[t]
=&\sqrt{\rho}\left(-\frac{12 \rho ^2 y^2}{(y-1)^2}-\frac{24 \rho\, y^2}{y-1}-12 y^2\right)\\
&+\sqrt{\rho}\left(\frac{4 (5-2 y) \rho ^2 y^3}{(y-1)^4}+\frac{4 (5-3 y) \rho\, y^3}{(y-1)^3}\right) \frac{\mu_\pi^2}{m_b^2}\\
&+\sqrt{\rho}\left(\frac{12 \rho\, y^3}{(y-1)^2}-\frac{36 (y-2) \rho ^2 y^2}{(y-1)^3}+\frac{24 (2 y-3) y^2}{y-1}\right) \frac{\mu_g^2}{3 \,m_b^2}
\end{aligned}\\ \\
\left(\frac{\text{d} \Gamma}{\text{d} y}\right)^{c_L g_L}&
\begin{aligned}[t]
=&\left(-\frac{12 \rho ^2  y^2}{y-1}-12 (y-1)  y^2-24 \rho\, y^2\right) m_b\\
&+\left(-\frac{2 \left(4 y^2-9 y+3\right) \rho ^2  y^2}{(y-1)^3}-12 \rho\, y^2+6  y^2\right) \frac{\mu_\pi^2}{m_b^2}\\
&+\left(\frac{6 (3-2 y) \rho ^2  y^2}{(y-1)^2}-\frac{12 (y-3) \rho\, y^2}{y-1}+18  y^2\right) \frac{\mu_g^2}{3 \,m_b^2}
\end{aligned}\\ \\
\end{align*}

\begin{align*}
\left(\frac{\text{d} \Gamma}{\text{d} y}\right)^{c_L g_R}&
\begin{aligned}[t]
=&\sqrt{\rho}\left(-\frac{12 \rho ^2  y^2}{y-1}-12 (y-1)  y^2-24 \rho\, y^2\right)m_b \\
&+\sqrt{\rho}\left(-\frac{2 \left(4 y^2-9 y+3\right) \rho ^2  y^2}{(y-1)^3}-12 \rho\, y^2+6  y^2\right) \frac{\mu_\pi^2}{m_b^2}\\
&+\sqrt{\rho}\left(\frac{30 (3-2 y) \rho ^2  y^2}{(y-1)^2}-\frac{60 (y-3) \rho\, y^2}{y-1}+90  y^2\right) \frac{\mu_g^2}{3 \,m_b^2}
\end{aligned}\\ \\
\left(\frac{\text{d} \Gamma}{\text{d} y}\right)^{c_L d_L}&
\begin{aligned}[t]
=&\left(-\frac{8 \rho ^3  y^3}{(y-1)^3}-\frac{12 \rho ^2  y^3}{(y-1)^2}+4  y^3\right)m_b \\
&+\left(\frac{4 \left(-4 y^2+11 y+5\right) \rho ^3  y^3}{3 (y-1)^5}+\frac{2 \left(-3 y^2+4 y+5\right) \rho ^2  y^3}{(y-1)^4}-\frac{10  y^3}{3}\right) \frac{\mu_\pi^2}{m_b^2}\\
&+\left(\frac{12 (5-2 y) \rho ^3  y^3}{(y-1)^4}-\frac{6 (y-3) \rho ^2  y^3}{(y-1)^3}+\frac{24 (y-2) \rho\, y^3}{(y-1)^2}+6  y^3\right) \frac{\mu_g^2}{3 \,m_b^2}
\end{aligned}\\ \\
\left(\frac{\text{d} \Gamma}{\text{d} y}\right)^{c_L d_R}&
\begin{aligned}[t]\nonumber
=&\sqrt{\rho}\left(\frac{12 (y-3) \rho ^2  y^2}{(y-1)^2}+4 (y-3)  y^2+\frac{12 (y-3) \rho\, y^2}{y-1}\right)m_b \\
&	\begin{aligned}[t] +\sqrt{\rho}\Biggl(
		\frac{8 \left(y^2-5 y+10\right) \rho ^3  y^3}{3 (y-1)^5}+\frac{2 \left(3 y^2-16 y+25\right) \rho ^2  y^3}{(y-1)^4}
		+\frac{4 (5-3 y) \rho\, y^3}{(y-1)^3}-\frac{10  y^3}{3}\Biggr) \frac{\mu_\pi^2}{m_b^2}
	\end{aligned}\\
&	\begin{aligned}[t] +\sqrt{\rho}\Biggl(
		&\frac{20 y^2 \left(y^2-4 y+6\right)  \rho ^3}{(y-1)^4}+\frac{6 y^2 \left(5 y^2-25 y+36\right)  \rho ^2}{(y-1)^3}\\
		&\hspace{2cm}+\frac{12 (6-5 y) y^2  \rho }{(y-1)^2}-\frac{2 y^2 \left(5 y^2-5 y+12\right) }{y-1}\Biggr) \frac{\mu_g^2}{3 \,m_b^2}
	\end{aligned}
\end{aligned}
\end{align*}

\vspace{1cm}

\begin{equation*}
\begin{split}
\frac{\text{d}\Gamma}{\text{d}y}=\frac{G_F^2 m_b^5}{192 \pi^3}\vert V_{cb}\vert
\Biggl(&
\left(\frac{\text{d} \Gamma}{\text{d} y}\right)^{c_L c_L}\,c_L^2 +
\left(\frac{\text{d} \Gamma}{\text{d} y}\right)^{c_L c_R}\,c_L\, c_R +
\left(\frac{\text{d} \Gamma}{\text{d} y}\right)^{c_L g_L}\,c_L\,g_L  \\
&+\left(\frac{\text{d} \Gamma}{\text{d} y}\right)^{c_L g_R}\,c_L\,g_R +
\left(\frac{\text{d} \Gamma}{\text{d} y}\right)^{c_L d_L}\,c_L\,d_L +
\left(\frac{\text{d} \Gamma}{\text{d} y}\right)^{c_L d_R}\,c_L\,d_R
\Biggr)
\end{split}
\end{equation*}

\newpage

\section{Tables}
\enlargethispage{1cm}
\begin{table}[!!h]
    \centering
    \begin{tabular}{cccc|d4d4d4d4d4d4}\toprule
        &&&\textbf{n} & \multicolumn{1}{c}{\boldmath$\;c_L^2$} & \multicolumn{1}{c}{\boldmath$\;c_Lc_R$} & \multicolumn{1}{c}{\boldmath$\;\;c_Lg_L$} & \multicolumn{1}{c}{\boldmath$\;\;c_Lg_R$} & \multicolumn{1}{c}{\boldmath$\;\;c_Ld_L$} & \multicolumn{1}{c}{\boldmath$\;c_Ld_R$} \\
        \midrule
        \multirow{16}{*}{\rotatebox{90}{no $E_l$ cuts}}&
        \multirow{12}{*}{\rotatebox{90}{\hspace{-0.5cm}Tree-level}}&
         \multirow{4}{*}{\rotatebox{90}{parton}}
         & 0 &   1.0000 &  -0.6685 &   0.2212 &   0.5400 &   0.3315 &  -0.6597 \\
        &&& 1 &   0.3072 &  -0.2092 &   0.0613 &   0.1372 &   0.0977 &  -0.2307 \\
        &&& 2 &   0.1030 &  -0.0708 &   0.0188 &   0.0388 &   0.0314 &  -0.0845 \\
        &&& 3 &   0.0365 &  -0.0252 &   0.0062 &   0.0118 &   0.0107 &  -0.0319 \\
        \cmidrule{3-10}
        &&
        \multirow{4}{*}{\rotatebox{90}{$\mu_\pi^2 / m_b^2$}}
         \multirow{4}{*}{\rotatebox{90}{ coeff.}}
         & 0 & -0.5000 & 0.3342 & -0.0017 & 0.1703 & -0.1652 & 0.3288 \\
        &&& 1 & 0.0000 & 0.0000 &  0.0000 & 0.0000 &   0.0000 & 0.0000 \\
        &&& 2 & 0.0858 & -0.0590 & 0.0365 & 0.1146 & 0.0261 & -0.0702 \\
        &&& 3 & 0.0730 & -0.0503 & 0.0210 & 0.0575 & 0.0214 & -0.0637\\
        \cmidrule{3-10}
        &&
        \multirow{4}{*}{\rotatebox{90}{$\mu_g^2 / m_b^2$}}
        \multirow{4}{*}{\rotatebox{90}{ coeff.}}
         & 0 & -1.9449 & 4.9934 & 1.0232 & 1.5624 & -2.1536 & 3.7106 \\
        &&& 1 & -0.9625 & 1.8578 & 0.3253 & 0.6011 & -0.7986 & 1.5873 \\
        &&& 2 & -0.4495 & 0.7237 & 0.1124 & 0.2427 & -0.3081 & 0.6840 \\
        &&& 3 & -0.2052 & 0.2902 & 0.0410 & 0.1008 & -0.1220 & 0.2966 \\
        \cmidrule{2-10}
         &\multirow{4}{*}{\rotatebox{90}{$\alpha_s/\pi$ coeff.}}
         && 0 &   0.3125 &   0.8009 &  -2.6592 &  -8.8212 &  -2.1497 &   4.3637 \\
        &&& 1 &   0.0908 &   0.2284 &  -0.7171 &  -2.3141 &  -0.5594 &   1.4880 \\
        &&& 2 &   0.0276 &   0.0739 &  -0.2174 &  -0.6843 &  -0.1660 &   0.5394 \\
        &&& 3 &   0.0085 &   0.0260 &  -0.0711 &  -0.2189 &  -0.0538 &   0.2039 \\
        \cmidrule{1-10}
        \multirow{16}{*}{\rotatebox{90}{$E_l>\unit[1]{GeV}$ cut}}
        &\multirow{12}{*}{\rotatebox{90}{\hspace{-0.5cm}Tree-level}}
         & \multirow{4}{*}{\rotatebox{90}{parton}}
         & 0 &   0.8148 &  -0.5617 &   0.1621 &   0.3586 &   0.2631 &  -0.6161 \\
        &&& 1 &   0.2776 &  -0.1919 &   0.0520 &   0.1089 &   0.0867 &  -0.2232 \\
        &&& 2 &   0.0979 &  -0.0678 &   0.0172 &   0.0340 &   0.0296 &  -0.0831 \\
        &&& 3 &   0.0356 &  -0.0246 &   0.0059 &   0.0109 &   0.0104 &  -0.0317 \\
        \cmidrule{3-10}
        && \multirow{4}{*}{\rotatebox{90}{$\mu_\pi^2 / m_b^2$ }}
        \multirow{4}{*}{\rotatebox{90}{ coeff.}}
         & 0 & -0.4504 & 0.3225 & 0.0433 & 0.3440 & -0.1479 & 0.3631 \\
        &&&  1 &0.0087 & -0.0021 & 0.0564 & 0.2247 & 0.0031 & 0.0059 \\
        &&&  2 & 0.0874 & -0.0594 & 0.0377 & 0.1194 & 0.0267 & -0.0691 \\
        &&&  3 &0.0733 & -0.0504 & 0.0213 & 0.0583 & 0.0215 & -0.0635 \\
        \cmidrule{3-10}
        && \multirow{4}{*}{\rotatebox{90}{$\mu_g^2 / m_b^2$}}
        \multirow{4}{*}{\rotatebox{90}{ coeff.}}
         & 0 & -2.1029 & 4.6903 & 0.8592 & 1.4595 & -2.0451 & 3.7102 \\
        &&& 1 & -0.9883 & 1.8078 & 0.2989 & 0.5845 & -0.7805 & 1.5871 \\
        &&& 2 & -0.4540 & 0.7149 & 0.1078 & 0.2398 & -0.3049 & 0.6840 \\
        &&& 3 & -0.2060 & 0.2886 & 0.0401 & 0.1003 & -0.1214 & 0.2966 \\
        \cmidrule{2-10}
        &\multirow{4}{*}{\rotatebox{90}{$\alpha_s/\pi$ coeff.}}
         && 0 &   0.2640 &   0.5740 &  -1.8506 &  -5.9374 &  -1.3992 &   3.9213 \\
        &&& 1 &   0.0828 &   0.1930 &  -0.5920 &  -1.8692 &  -0.4440 &   1.4126 \\
        &&& 2 &   0.0262 &   0.0679 &  -0.1964 &  -0.6098 &  -0.1467 &   0.5260 \\
        &&& 3 &   0.0083 &   0.0249 &  -0.0674 &  -0.2058 &  -0.0504 &   0.2014 \\
        \bottomrule
    \end{tabular}
    \caption{\label{tab:L}Tree level and $\alpha_s/\pi$ coefficients of the leptonic moments without $E_l$ cuts and with a cut  $E_l{\,>\,}\unit[1]{GeV}$. Note that we have redefined ${\bf d_{L/R} } = m_B\,d_{L/R}$ and
  ${\bf g_{L/R} } = m_B\,g_{L/R}$ with $m_B=\unit[5.279]{GeV}$ in order to tabulate dimensionless quantities. }
\end{table}

\begin{table}[!!h]
    \centering
    \begin{tabular}{cccc|d4d4d4d4d4d4}\toprule
        &&\textbf{i}&\textbf{j}& \multicolumn{1}{c}{\boldmath$c_L^2$} & \multicolumn{1}{c}{\boldmath$c_Lc_R$} & \multicolumn{1}{c}{\boldmath$c_Lg_L$} & \multicolumn{1}{c}{\boldmath$c_Lg_R$} & \multicolumn{1}{c}{\boldmath$c_Ld_L$} & \multicolumn{1}{c}{\boldmath$c_Ld_R$} \\
        \midrule
        \multirow{23}{*}{\rotatebox{90}{\hspace{1.25cm}Tree-level}}
        &\multirow{5}{*}{\rotatebox{90}{parton}}
        & 0 & 0 &   1.0000 &  -0.6685 &   0.2212 &   0.5400 &   0.3315 &  -0.6597 \\
        && 0 & 1 &   0.4220 &  -0.2500 &   0.0961 &   0.2556 &   0.1217 &  -0.2559 \\
        && 0 & 2 &   0.1832 &  -0.0964 &   0.0429 &   0.1219 &   0.0461 &  -0.1021 \\
        && 0 & 3 &   0.0815 &  -0.0383 &   0.0196 &   0.0586 &   0.0180 &  -0.0418 \\
        && $i > 0$ & $j$ & 0.0000  & 0.0000  & 0.0000  & 0.0000  & 0.0000  & 0.0000\\
        \cmidrule{2-10}
        &\multirow{10}{*}{\rotatebox{90}{$\mu_\pi^2 / m_b^2$ coeff.}}
        & 0 & 0 & -0.5000 & 0.3342 & -0.0017 & 0.1703 & -0.1652 & 0.3288 \\
        && 0 & 1 & -0.5000 & 0.3342 & -0.100 & -0.2229 & -0.1652 & 0.3288 \\
        && 0 & 2 & -0.2902 & 0.1836 & -0.0773 & -0.2119 & -0.0899 & 0.1840 \\
        && 0 & 3 & -0.1382 & 0.0837 & -0.0448 & -0.1348 & -0.0406 & 0.0853 \\
        && 1 & 0 & -0.5780 & 0.4185 & -0.2038 & -0.5937 & -0.2091 & 0.4025 \\
        && 1 & 1 & -0.1584 & 0.1172 & -0.0695 & -0.2158 & -0.0585 & 0.1129 \\
        && 1 & 2 & -0.0283 & 0.0280 & -0.0217 & -0.0718 & -0.0143 & 0.0258 \\
        && 2 & 0 & 0.1609 & -0.0728 & 0.0386 & 0.1159 & 0.0337 & -0.0809 \\
        && 2 & 1 &   0.0735 & -0.0302 & 0.0180 & 0.0561 & 0.0138 & -0.0343 \\
        && 3 & 0 &   0.0000 & 0.0000 & 0.0000 & 0.0000 & 0.0000 & -0.0000 \\
        \cmidrule{2-10}
        &\multirow{8}{*}{\rotatebox{90}{$\mu_g^2 / m_b^2$ coeff.}}
        & 0 & 0 & -1.9449 & 4.9934 & 1.0232 & 1.5624 & -2.1536 & 3.7106 \\
        && 0 & 1 & -0.3850 & 1.2777 & 0.4097 & 0.4782 & -0.5223 & 0.9700 \\
        && 0 & 2 & -0.0302 & 0.2833 & 0.1576 & 0.1391 & -0.1109 & 0.2254 \\
        && 0 & 3 &  0.0298 & 0.0342 & 0.0578 & 0.0350 & -0.0146 & 0.0347 \\
        && 1 & 0 &  0.3143 & -0.6395 & -0.1100 & -0.2167 & 0.2027 & -0.4360 \\
        && 1 & 1 &  0.1195 & -0.2561 & -0.0529 & -0.0925 & 0.0744 & -0.1709 \\
        && 1 & 2 &  0.0466 & -0.1059 & -0.0254 & -0.0405 & 0.0282 & -0.0689 \\
        && $i>1$ & $j$ &  0.0000 & 0.0000 & 0.0000 & 0.0000 & 0.0000 & 0.0000 \\
        \midrule
        \multirow{10}{*}{\rotatebox{90}{$\alpha_s/\pi$ coeff.}}
        && 0 & 0 &   0.3128 &   0.8007 &  -2.6592 &  -8.8212 &  -2.1497 &   4.3637 \\
        && 0 & 1 &   0.1631 &   0.3441 &  -1.2391 &  -4.1901 &  -0.8839 &   1.8575 \\
        && 0 & 2 &   0.0910 &   0.1477 &  -0.5850 &  -2.0067 &  -0.3694 &   0.8017 \\
        && 0 & 3 &   0.0526 &   0.0632 &  -0.2793 &  -0.9681 &  -0.1568 &   0.3505 \\
        && 1 & 0 &   0.0901 &  -0.0363 &   0.0028 &   0.0176 &   0.0032 &  -0.0095 \\
        && 1 & 1 &   0.0470 &  -0.0178 &   0.0014 &   0.0093 &   0.0015 &  -0.0046 \\
        && 1 & 2 &   0.0251 &  -0.0090 &   0.0007 &   0.0050 &   0.0007 &  -0.0023 \\
        && 2 & 0 &   0.0091 &  -0.0033 &   0.0001 &   0.0015 &   0.0002 &  -0.0008 \\
        && 2 & 1 &   0.0053 &  -0.0019 &   0.0000 &   0.0009 &   0.0001 &  -0.0004 \\
        && 3 & 0 &   0.0018 &  -0.0006 &   0.0000 &   0.0003 &   0.0000 &  -0.0001 \\
        \bottomrule
    \end{tabular}
    \caption{\label{tab:Hnocut}Tree level and $\alpha_s/\pi$ coefficients of the hadronic moments without $E_l$ cuts. The partonic tree-level moments for $i>1$ are all zero. Note that we have redefined
    ${\bf d_{L/R} } = m_B\,d_{L/R}$ and ${\bf g_{L/R} } = m_B\,g_{L/R}$ with $m_B=\unit[5.279]{GeV}$ in order to tabulate dimensionless quantities. }
\end{table}

\begin{table}[!!h]
    \centering
    \begin{tabular}{cccc|d4d4d4d4d4d4}\toprule
        &&\textbf{i}&\textbf{j}& \multicolumn{1}{c}{\boldmath$c_L^2$} & \multicolumn{1}{c}{\boldmath$c_Lc_R$} & \multicolumn{1}{c}{\boldmath$c_Lg_L$} & \multicolumn{1}{c}{\boldmath$c_Lg_R$} & \multicolumn{1}{c}{\boldmath$c_Ld_L$} & \multicolumn{1}{c}{\boldmath$c_Ld_R$} \\
        \midrule
        \multirow{23}{*}{\rotatebox{90}{\hspace{1.25cm}Tree-level}}
        &\multirow{5}{*}{\rotatebox{90}{parton}}
        & 0 & 0 &   0.8148 &  -0.5617 &   0.1621 &   0.3586 &   0.2631 &  -0.6161 \\
        && 0 & 1 &   0.3341 &  -0.2037 &   0.0682 &   0.1676 &   0.0922 &  -0.2365 \\
        && 0 & 2 &   0.1411 &  -0.0761 &   0.0295 &   0.0789 &   0.0332 &  -0.0933 \\
        && 0 & 3 &   0.0612 &  -0.0293 &   0.0131 &   0.0375 &   0.0123 &  -0.0378 \\
        && $i > 0$ & $j$ & 0.0000  & 0.0000  & 0.0000  & 0.0000  & 0.0000  & 0.0000\\
        \cmidrule{2-10}
        &\multirow{10}{*}{\rotatebox{90}{$\mu_\pi^2 / m_b^2$ coeff.}}
        & 0 & 0 & -0.4504 & 0.3225 & 0.0433 & 0.3440 & -0.1479 & 0.3631 \\
        && 0 & 1 & -0.4505 & 0.2921 & -0.0597 & -0.0843 & -0.1329 & 0.3332 \\
        && 0 & 2 & -0.2673 & 0.1561 & -0.0532 & -0.1300 & -0.0695 & 0.1841 \\
        && 0 & 3 & -0.1337 & 0.0706 & -0.0327 & -0.0935 & -0.0308 & 0.0859 \\
        && 1 & 0 & -0.5424 & 0.3590 & -0.1687 & -0.4845 & -0.1685 & 0.3887 \\
        && 1 & 1 & -0.1639 & 0.1022 & -0.0598 & -0.1852 & -0.0478 & 0.1115 \\
        && 1 & 2 & -0.0417 & 0.0262 & -0.0204 & -0.0678 & -0.0126 & 0.0273 \\
        && 2 & 0 &  0.1203 & -0.0547 & 0.0258 & 0.0742 & 0.0223 & -0.0729 \\
        && 2 & 1 &  0.0538 & -0.0221 & 0.0118 & 0.0355 & 0.0087 & -0.0306 \\
        && 3 & 0 &  0.0000 &  0.0000 &   0.0000 &   0.0000 &   0.0000 &  0.0000 \\
        \cmidrule{2-10}
        &\multirow{8}{*}{\rotatebox{90}{$\mu_g^2 / m_b^2$ coeff.}}
        & 0 & 0 & -2.1029 & 4.6903 & 0.8592 & 1.4595 & -2.0451 & 3.7102 \\
        && 0 & 1 & -0.4609 & 1.2205 & 0.3461 & 0.4476 & -0.5005 & 0.9855 \\
        && 0 & 2 & -0.0660 & 0.2921 & 0.1348 & 0.1332 & -0.1119 & 0.2391 \\
        && 0 & 3 &  0.0131 & 0.0538 & 0.0507 & 0.0363 & -0.0194 & 0.0439 \\
        && 1 & 0 &  0.3074 & -0.5095 & -0.0803 & -0.1804 & 0.1654 & -0.4093 \\
        && 1 & 1 &  0.1171 & -0.1971 & -0.0381 & -0.0751 & 0.0583 & -0.1590 \\
        && 1 & 2 &  0.0458 & -0.0789 & -0.0180 & -0.0321 & 0.0211 & -0.0635 \\
        && $i > 1$ & $j$ &  0.0000 & 0.0000 & 0.0000 & 0.0000 & 0.0000 & 0.0000 \\
        \midrule
        \multirow{10}{*}{\rotatebox{90}{$\alpha_s/\pi$ coeff.}}
        && 0 & 0 &   0.2642 &   0.5739 &  -1.8506 &  -5.9373 &  -1.3992 &   3.9213 \\
        && 0 & 1 &   0.1216 &   0.2462 &  -0.8449 &  -2.7806 &  -0.5529 &   1.6572 \\
        && 0 & 2 &   0.0608 &   0.1057 &  -0.3919 &  -1.3149 &  -0.2221 &   0.7103 \\
        && 0 & 3 &   0.0323 &   0.0455 &  -0.1842 &  -0.6272 &  -0.0907 &   0.3086 \\
        && 1 & 0 &   0.0576 &  -0.0231 &   0.0018 &   0.0101 &   0.0018 &  -0.0079 \\
        && 1 & 1 &   0.0288 &  -0.0108 &   0.0009 &   0.0052 &   0.0008 &  -0.0038 \\
        && 1 & 2 &   0.0147 &  -0.0052 &   0.0004 &   0.0027 &   0.0004 &  -0.0018 \\
        && 2 & 0 &   0.0046 &  -0.0016 &   0.0001 &   0.0007 &   0.0001 &  -0.0006 \\
        && 2 & 1 &   0.0026 &  -0.0009 &   0.0000 &   0.0004 &   0.0000 &  -0.0003 \\
        && 3 & 0 &   0.0007 &  -0.0002 &   0.0000 &   0.0001 &   0.0000 &  -0.0001 \\
        \bottomrule
    \end{tabular}
    \caption{\label{tab:Hcut}Tree level and $\alpha_s/\pi$ coefficients of the hadronic moments  with a cut  $E_l{\,>\,}\unit[1]{GeV}$. The partonic tree-level moments for $i>1$ are all zero. Note that we have redefined
    ${\bf d_{L/R} } = m_B\,d_{L/R}$ and ${\bf g_{L/R} } = m_B\,g_{L/R}$ with $m_B=\unit[5.279]{GeV}$ in order to tabulate dimensionless quantities. }
\end{table}

\begin{table}[!!h]
    \centering
    \begin{tabular}{cc|d4d4d4d4d4d4}\toprule
        {\boldmath$\mu$}&\textbf{n} & \multicolumn{1}{c}{\boldmath$\;c_L^2$} & \multicolumn{1}{c}{\boldmath$\;c_Lc_R$} & \multicolumn{1}{c}{\boldmath$\;\;c_Lg_L$} & \multicolumn{1}{c}{\boldmath$\;\;c_Lg_R$} & \multicolumn{1}{c}{\boldmath$\;\;c_Ld_L$} & \multicolumn{1}{c}{\boldmath$\;c_Ld_R$} \\
        \midrule
        \multirow{4}{*}{\rotatebox{90}{$\unit[2.3]{GeV}$}}
        & 0 &   1.0253 &  -0.6037 &   0.0042 &  -0.1916 &   0.1533 &  -0.2983 \\
        & 1 &   0.3145 &  -0.1907 &   0.0028 &  -0.0552 &   0.0512 &  -0.1074 \\
        & 2 &   0.1052 &  -0.0648 &   0.0011 &  -0.0182 &   0.0176 &  -0.0397 \\
        & 3 &   0.0372 &  -0.0231 &   0.0004 &  -0.0065 &   0.0062 &  -0.0150 \\
        \midrule
        \multirow{4}{*}{\rotatebox{90}{$\unit[4.6]{GeV}$}}
        & 0 &   1.0208 &  -0.6151 &   0.0441 &  -0.0474 &   0.1883 &  -0.3692 \\
        & 1 &   0.3132 &  -0.1940 &   0.0135 &  -0.0169 &   0.0604 &  -0.1317 \\
        & 2 &   0.1048 &  -0.0658 &   0.0043 &  -0.0068 &   0.0204 &  -0.0485 \\
        & 3 &   0.0371 &  -0.0234 &   0.0015 &  -0.0028 &   0.0071 &  -0.0184 \\
        \midrule
        \multirow{4}{*}{\rotatebox{90}{$\unit[9.2]{GeV}$}}
        & 0 &   1.0177 &  -0.6231 &   0.0715 &   0.0752 &   0.2146 &  -0.4223 \\
        & 1 &   0.3123 &  -0.1963 &   0.0208 &   0.0164 &   0.0674 &  -0.1499 \\
        & 2 &   0.1046 &  -0.0666 &   0.0065 &   0.0034 &   0.0225 &  -0.0552 \\
        & 3 &   0.0370 &  -0.0237 &   0.0022 &   0.0005 &   0.0078 &  -0.0209 \\
        \bottomrule
    \end{tabular}
    \caption{\label{tab:LSummed}Summed up tree level and $\alpha_s/\pi$ coefficients of the leptonic moments without $E_l$ cuts for $\mu=2.3, 4.6$ and $\unit[9.2]{GeV}$.}
    \vspace{1.5cm}
\end{table}

\begin{table}[!!h]
    \centering
    \begin{tabular}{ccc|d4d4d4d4d4d4}\toprule
        {\boldmath$\mu$}&\textbf{i}&\textbf{j} & \multicolumn{1}{c}{\boldmath$\;c_L^2$} & \multicolumn{1}{c}{\boldmath$\;c_Lc_R$} & \multicolumn{1}{c}{\boldmath$\;\;c_Lg_L$} & \multicolumn{1}{c}{\boldmath$\;\;c_Lg_R$} & \multicolumn{1}{c}{\boldmath$\;\;c_Ld_L$} & \multicolumn{1}{c}{\boldmath$\;c_Ld_R$} \\
        \midrule
        \multirow{4}{*}{\rotatebox{90}{$\unit[2.3]{GeV}$}}
        & 0 & 0 &   1.0253 &  -0.6037 &   0.0042 &  -0.1916 &   0.1533 &  -0.2983 \\
        & 0 & 1 &   0.4352 &  -0.2222 &  -0.0051 &  -0.0914 &   0.0486 &  -0.1024 \\
        & 0 & 2 &   0.1906 &  -0.0845 &  -0.0049 &  -0.0440 &   0.0156 &  -0.0360 \\
        & 0 & 3 &   0.0857 &  -0.0331 &  -0.0033 &  -0.0214 &   0.0050 &  -0.0129 \\
        \midrule
        \multirow{4}{*}{\rotatebox{90}{$\unit[4.6]{GeV}$}}
        & 0 & 0 &   1.0208 &  -0.6151 &   0.0441 &  -0.0474 &   0.1883 &  -0.3692 \\
        & 0 & 1 &   0.4329 &  -0.2271 &   0.0136 &  -0.0234 &   0.0628 &  -0.1322 \\
        & 0 & 2 &   0.1892 &  -0.0866 &   0.0040 &  -0.0117 &   0.0215 &  -0.0487 \\
        & 0 & 3 &   0.0850 &  -0.0340 &   0.0010 &  -0.0059 &   0.0075 &  -0.0185 \\
        \midrule
        \multirow{4}{*}{\rotatebox{90}{$\unit[9.2]{GeV}$}}
        & 0 & 0 &   1.0177 &  -0.6231 &   0.0715 &   0.0752 &   0.2146 &  -0.4223 \\
        & 0 & 1 &   0.4312 &  -0.2305 &   0.0269 &   0.0335 &   0.0734 &  -0.1545 \\
        & 0 & 2 &   0.1883 &  -0.0880 &   0.0104 &   0.0151 &   0.0258 &  -0.0582 \\
        & 0 & 3 &   0.0845 &  -0.0347 &   0.0041 &   0.0069 &   0.0093 &  -0.0226 \\
        \bottomrule
    \end{tabular}
    \caption{\label{tab:HSummed}Summed up tree level and $\alpha_s/\pi$ coefficients of the non-zero-tree-level hadronic moments without $E_l$ cuts for $\mu=2.3, 4.6$ and $\unit[9.2]{GeV}$.}
\vspace{1cm}
\end{table}

\end{document}